\documentclass{article}

\usepackage[preprint]{neurips_2024}

\usepackage[utf8]{inputenc} 
\usepackage[T1]{fontenc}    
\usepackage{hyperref}       
\usepackage{url}            
\usepackage{booktabs}       
\usepackage{amsfonts}       
\usepackage{nicefrac}       
\usepackage{microtype}      
\usepackage{xcolor}         
\usepackage{multirow}
\usepackage{graphicx}
\usepackage{float}
\usepackage[numbers]{natbib}

\title{Explainable AI Enhances Glaucoma Referrals, Yet the Human-AI Team Still Falls Short of the AI Alone}

\author{%
  Catalina Gomez\thanks{joint first authors, allowed to change author order in their respective CVs} \\
  Department of Computer Science\\
  Johns Hopkins University\\
  \And
  Ruolin Wang$^\ast$\\
  Department of Computer Science\\
  Johns Hopkins University\\
  \And
  Katharina Breininger \\
  Department of Biomedical Engineering \\
  Friedrich-Alexander-Universität \\
  \And
  Corinne Casey\\
  Wilmer Eye Institute \\
  Johns Hopkins University School of Medicine\\
  \And
  Chris Bradley\\
  Wilmer Eye Institute \\
  Johns Hopkins University School of Medicine\\
  \And
  Mitchell Pavlak\\
  Department of Computer Science\\
  Johns Hopkins University\\
    \And
  Alex Pham\\
   Johns Hopkins University School of Medicine\\
  \And
  Jithin Yohannan$^\dagger$\\
  Wilmer Eye Institute \\
  Johns Hopkins University School of Medicine\\
    \And
  Mathias Unberath \thanks{joint senior authors, allowed to change author order in thier respective CVs}\\
  Department of Computer Science\\
  Johns Hopkins University\\
}

\begin{document}

\maketitle

\begin{abstract}
Primary care providers are vital for initial triage and referrals to specialty care. In glaucoma, asymptomatic and fast progression can lead to vision loss, necessitating timely referrals to specialists. However, primary eye care providers may not identify urgent cases, potentially delaying care. Artificial Intelligence (AI) offering explanations could enhance their referral decisions. We investigate how various AI explanations help providers distinguish between patients needing immediate or non-urgent specialist referrals.
We built explainable AI algorithms to predict glaucoma surgery needs from routine eyecare data as a proxy for identifying high-risk patients. We incorporated intrinsic and post-hoc explainability and conducted an online study with optometrists to assess human-AI team performance, measuring referral accuracy and analyzing interactions with AI, including agreement rates, task time, and user experience perceptions.
AI support enhanced referral accuracy among 87 participants (59.9\%/50.8\% with/without AI), though Human-AI teams underperformed compared to AI alone. Participants believed they included AI advice more when using the intrinsic model, and perceived it more useful and promising. Without explanations, deviations from AI recommendations increased. AI support did not increase workload, confidence, and trust, but reduced challenges. On a separate test set, our black-box and intrinsic models achieved an accuracy of 77\% and 71\%, respectively, in predicting surgical outcomes. 
We identify opportunities of human-AI teaming for glaucoma management in primary eye care, noting that while AI enhances referral accuracy, it also shows a performance gap compared to AI alone, even with explanations. Human involvement remains essential in medical decision making, underscoring the need for future research to optimize collaboration, ensuring positive experiences and safe AI use.
\end{abstract}

\section{Introduction}
Primary care providers, serving as the first point of contact within the healthcare system and trained to manage a wide range of health issues, will refer patients to medical specialists when their conditions require more advanced treatment or diagnosis. The time frame from recognizing a health issue to reaching specialty care can vary depending on the severity and urgency of the conditions, and ultimately specialist availability. Timing is crucial in the early detection of asymptomatic conditions with irreversible damage like glaucoma to prevent severe outcomes such as vision loss, necessitating prompt referral of high-risk patients to glaucoma specialists for more aggressive intervention~\cite{stein2021glaucoma}. Proper care for glaucoma patients is essential, particularly as the disease can progress rapidly or unnoticed, with some patients remaining asymptomatic even until late in the disease course~\cite{tan2020glaucoma}. Therefore, from the initial consultations with primary eye care providers, it becomes imperative to distinguish patients requiring immediate glaucoma treatment and define longer follow-up intervals for patients at lower risk.

This initial step is crucial for navigating the complex clinical workflow. Glaucoma assessment and monitoring of disease progression include measurements of intraocular pressure (IOP), visual function via  visual field (VF) tests, and changes to the optic nerve head via optical coherence tomography (OCT)~\cite{stein2021glaucoma}. Glaucoma specialists  must individually tailor treatment decisions for each patient, taking into account the extent of IOP reduction needed to mitigate progression risk, encompassing options like medications, laser procedures, and incisional surgery~\cite{stein2021glaucoma}. Given this spectrum of interventions, the decision to opt for surgery, clearly documented in electronic health records (EHR), emerges as a pivotal indicator of high-risk glaucoma~\cite{wang2022deep,baxter2019machine}. Specifically, patients who require surgery within three months are often those at a greater risk of rapid disease progression, compared to those that undergo surgery further in the future. While glaucoma specialists have the expertise to make such critical decisions, primary eye care providers may not have the specialized training required for identifying patients at high-risk and make referral recommendations based on their urgency. Expanding the reach of glaucoma screening becomes increasingly critical in light of the expected growth in glaucoma patients, rendering it impractical for specialists alone to triage all glaucoma eyes and identify those at higher risk~\cite{resnikoff2020estimated}. Considering the importance of diagnosis and treating glaucoma early to slow its progression and prevent the severe consequences of delayed intervention, artificial intelligence (AI) may serve as a key enabler to empower primary eyecare providers to make informed referral decisions. In contrast to previously studied AI-based decision support systems, such as skin lesions diagnosis~\cite{combalia2022validation,tschandl2020human}, it is unclear whether AI can similarly assist first contact providers in making referral (rather than diagnostic) decisions more accurately. Moreover, questions persist about whether Explainable AI (XAI) provides a significant advantage over black-box AI models and how to tailor the information presented by XAI to best support decision-making in different clinical contexts. 

AI techniques for predictive modeling  are capable of analyzing vast datasets to assist in decisions such as target IOP level adjustments, visual field loss evaluation~\cite{wang2019artificial,herbert2023forecasting,shon2022can}, surgical intervention~\cite{wang2024deep,baxter2019machine} and patient progress monitoring~\cite{wang2022deep}, Such techniques offer promising avenues to enhance glaucoma care. Despite a long history of technological advancement with AI to risk stratify glaucoma, less is understood about how to appropriately present the results of these models to clinicians at the time they make decisions for individual patients. Equally important is gaining insight into how clinicians integrate AI recommendations into their clinical decision making. Particularly, in the glaucoma context, additional research is needed to understand the glaucoma clinical workflow and decision-making process for a successful integration of clinical decision support and its possibility to improve glaucoma outcomes~\cite{stagg2021special}. This research must delve into the transparency and interpretability of AI systems for clinician trust and adoption, not only demonstrating technical feasibility but also considering end users in the design and evaluation of human factors~\cite{chen2022explainable}. Recent empirical studies have explored explainability techniques within specific contexts, primarily focusing on experts accustomed to the clinical tasks being augmented by AI~\cite{naiseh2023different,tschandl2020human,panigutti2022understanding}. Conversely, research involving non-experts reveals potential biases in decision-making, including a tendency for overreliance on AI recommendations~\cite{gaube2023non}. Understanding the biases and challenges that arise for target users without task knowledge, and how AI guidance can address these, is crucial for enhancing healthcare agents' capabilities and patient care.

Recognizing this, we envision a human-AI teaming experience for providers to integrate AI assistance in patient referral decision making through transparent mechanisms. In this study, we develop a suite of explainable AI algorithms focused on triage recommendations, leveraging surgery predictions as a measure of urgency for specialist consultation. Using an in-house dataset of over 2,500 eyes and a user study with 87 primary eye care providers, we explored how AI support influences human behavior and perception in determining the urgency of referrals to glaucoma specialty care—immediately, in the near future, or not at all—based on ophthalmic patient vignettes. In particular, we tested the effect of AI-based systems and their explanations of different nature (post-hoc vs. intrinsic) on participants’ performance and experience during the referral decision making process. To comprehensively evaluate our algorithms, we calculated classification-related metrics in a larger test set and found that they delivered competitive performance without sacrificing accuracy for interpretability.

The human-teaming evaluation revealed that participants’ overall accuracy of referrals increased significantly compared to the non-assisted group, suggesting that AI had considerable impact on final recommendations but did not surpass the algorithmic performance levels. We measured two errors relevant for referral tasks, namely under and over-claims of the urgency level for each case. Assisted by the intrinsically interpretable AI, participants significantly reduced over-claiming the urgency of referrals compared to the non-assisted group, while no differences were found in under-claims. While participants reported varying levels of use of AI recommendations from different model types—with greater use reported for intrinsically interpretable models—the observed agreement rate was considerable (greater than 76\%), but did not change with the presence or type of explainability incorporated, aligned with trust findings. We quantify deviations in participants’ responses with respect to AI predictions considering differences in referral timing and found significantly more adjustments when the AI support did not include explanations. Subjectively, providers reported no increase in workload or confidence, but experienced significantly lower levels of challenge and frustration when using AI support for triaging tasks. In addition, perceived helpfulness and use in clinical practice was better for the intrinsically explainable model compared to the black box. Our work highlights the importance of empirical evaluations with human subjects to gauge the AI system's impact on user behavior and key outcomes in clinical practice and in this way improve algorithms and leverage AI to enhance human capabilities safely.

\section{Methods}
\paragraph{Hypotheses}
We explored the impact of AI assistance on referral decision-making within the context of glaucoma, focusing on how it affects the user experience and decision outcomes when interacting with AI systems that provide transparent reasoning. In particular, we formulate the following hypotheses: 

H1.a: Groups using AI-based systems exhibit better performance compared to groups that do not use AI-based systems.  

H1.b: Groups with AI explanations demonstrate better performance compared to groups using AI-based systems without explanations.

H2.a: Groups with AI support have a better decision-making experience compared to groups that do not use AI-based systems. 

H2.b: Groups with explanations have a better decision-making experience compared to groups using AI-based systems without explanations.

\begin{figure}[h]
    \centering
    \includegraphics[width=0.95\linewidth]{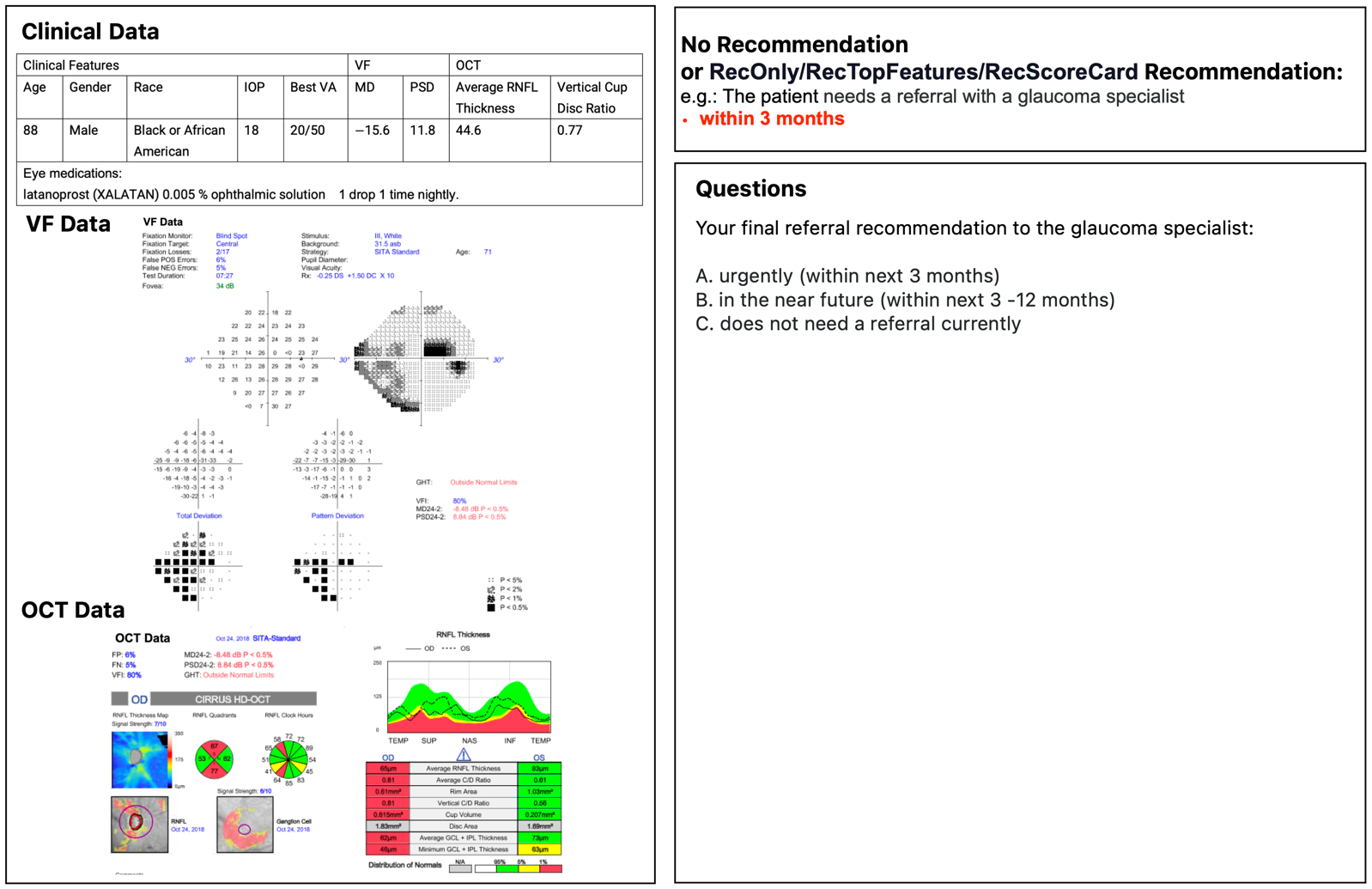}
    \caption{Visualization of the web interface for referral of patients to specialists. Participants first review the VF, OCT, and clinical data on the left side. The Human Only or AI recommendation (No Explanation/Feature Importance Explanations/Scoring-based Explanations) is displayed on the top right side. If present, explanations are displayed below the box that displays the AI recommendation (not shown in this image). Participants then decide if the patient needs a referral to glaucoma specialist within 3 months, 3-12 months, or does not need a referral currently.}
    \label{fig:interface}
\end{figure}

\paragraph{Human-AI task description and interface}
To evaluate the effectiveness of an AI-based referral system and providing transparency on its reasoning, we formulate a user study in which primary eye care provides (optometrists)  were asked to make referral recommendations to decide if the patient presented in a clinical vignette needs a referral to a glaucoma specialist within 3 months, 3-12 months, or does not need a referral currently, as illustrated in Figure~\ref{fig:interface}. The clinical vignette includes clinical features, such as age, gender, race, IOP, and best visual acuity (VA), outcomes of the ophthalmic tests, and eye medications. The VF test includes the mean deviation (MD) and pattern standard deviation (PSD), and visualizations of total deviation, pattern deviation and percentile plots. The report of the OCT test contains the average retinal nerve fiber layer (RNFL) thickness and vertical cup disc ratio, and visualizations of RNFL profile as well as  quadrant and clock hour maps. 

\paragraph{Conditions}
We designed four conditions that varied in whether and how an AI-based system assisted the participants in making patient referral decisions.

\begin{figure}[h]
    \centering
    \includegraphics[width=0.90\linewidth]{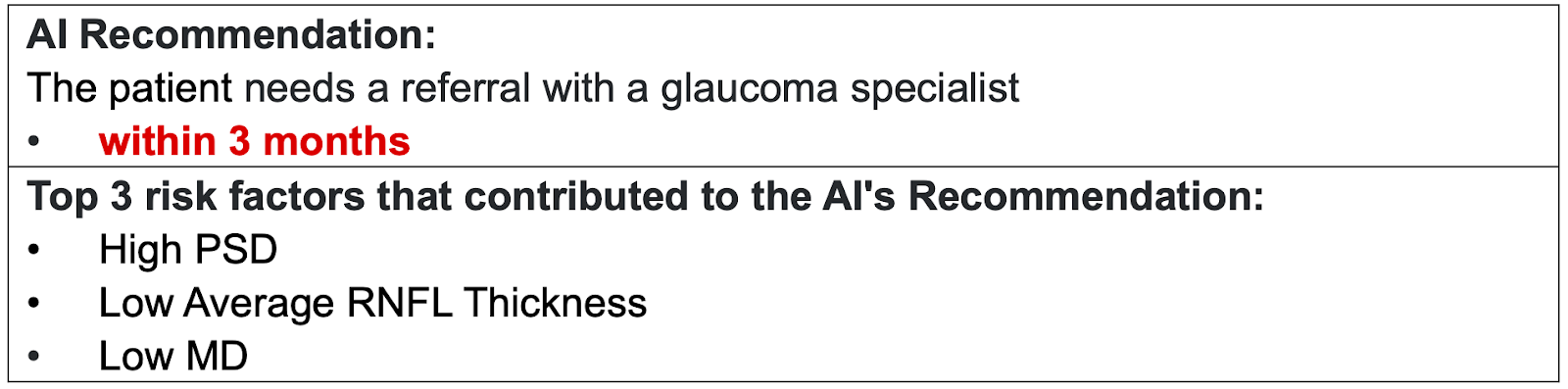}
    \caption{Example of Feature Importance Explanations. The AI recommendation is accompanied by the top three most important features calculated through SHAP values. The patient is recommended to schedule an urgent referral to a glaucoma specialist (within 3 months). High PSD, low average RNFL thickness, and low MD are identified as the top three most crucial features influencing the AI recommendation.}
    \label{fig:features}
\end{figure}

\begin{figure}[h]
    \centering
    \includegraphics[width=0.90\linewidth]{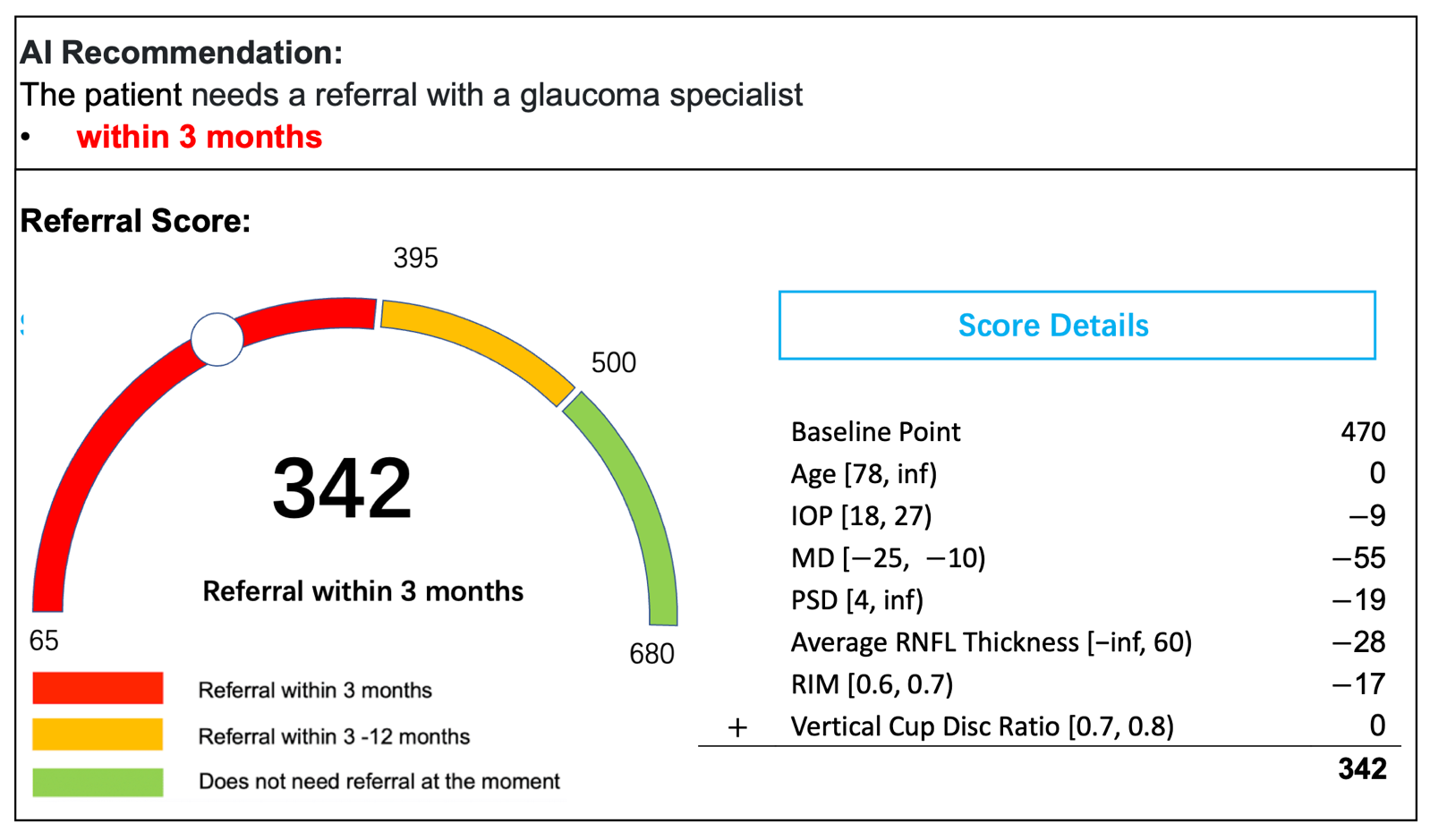}
    \caption{Example of Scoring-based Explanations. The AI recommendation is displayed with risk scores calculated using the Credit Scorecard method. The referral score is presented within a range that is associated with a referral recommendation, as shown on the left. Detailed rules for score calculation based on the feature distribution are provided on the right. The glaucoma risk scoring system for patient referral recommendation assigns a score to each risk factor ultimately integrated into the predictive model. Point scores calculation: The base score is set at 470, and the final score of 342 (derived from 470-9-55-19-28-17=342) is determined by the mathematical formula and corresponds to an urgent referral within 3 months.}
    \label{fig:scoring}
\end{figure}

\begin{itemize}
    \item Human Only: In the Human Only condition, participants were shown only the clinical data, VF data and OCT data of the patient. This condition served as a baseline where participants had to make decisions without any additional guidance or recommendations.

    \item No Explanation: In the No Explanation condition, participants were shown the clinical data, VF data, and OCT data of the patient, with the AI recommendation from a deep learning model (DLM) displayed and no additional information provided. Figure~\ref{fig:interface} illustrates how the AI prediction is displayed without any further information from the model. 

    \item Feature Importance Explanations: In the Feature Importance Explanations condition, participants were shown the clinical data, VF data, and OCT data of the patient. The AI recommendation from a DLM was displayed along with the top three most important features for an individual prediction, which were calculated using Shapley Additive Explanations (SHAP). SHAP values indicate each feature’s contribution to the prediction compared to a baseline, providing a local method to explain a prediction based on a single input patient case~\cite{lundberg2017unified}. As shown in Figure~\ref{fig:features}, high PSD, low average RNFL thickness, and low MD are identified as the top three most important features influencing the AI recommendation for a specific patient. For simplicity, we only display the feature names and not the magnitude of their contribution to the prediction. 

    \item Scoring-based Explanations: In the Scoring-based Explanations condition, participants were shown the clinical data, VF data, and OCT data of the patient. The AI recommendation was displayed with a visual risk scorecard and a transparent mathematical formula for determining a final score referral score. The coefficients for each feature in the formula were derived from a logistic regression combined with the weight-of-evidence (WoE), resulting in specific scores for each feature on the scorecard. This WoE process aims to normalize the feature scales and guarantee a steady and monotonic connection with the dependent variable. Based on the referral score, we define three regions that correspond to a different referral recommendation. In the Red region (score < 395), the patient is advised to schedule a referral to a glaucoma specialist within 3 months. In the Yellow region (395 < score < 500), the patient requires a referral to a glaucoma specialist within 3 - 12 months. In the Green Region (500 < score < 680), the patient does not currently require a referral. Figure~\ref{fig:scoring} displays the point scores for the Scoring-based Explanations that support the AI recommendation.

\end{itemize}

\paragraph{Models}
A DLM for surgery prediction and the logistic regression based credit scorecard model were utilized to provide referral recommendations. We utilized a vision transformer (ViT)~\cite{dosovitskiy2020image} in the DLM to extract features from visual representations of the VF and OCT data. Additionally, the input for the final classification model included numeric VF and OCT measures, along with clinical and demographic characteristics. The initial output was the probability of glaucoma surgery, which was then mapped to the three time horizon categories using a threshold that maximizes the surgical prediction accuracy in the validation set. Similarly, in the scorecard, we defined the region using a threshold derived from the score that optimizes accuracy. By selecting the threshold that achieved the highest accuracy on the validation set, we ensured that the region defined by this threshold maximized the model’s predictive performance. This approach allowed us to effectively segment the data based on the score, enabling more precise analysis and decision-making within the identified region. The Scorecard was created using the scorecard method with the scorecardpy Python package. We quantified the model’s performance using classification metrics, including accuracy, precision, and recall. We use the data collection from our initial work and follow the same data partitions to train and validate the models~\cite{wang2024deep}. To map the AI models outcomes to the recommendations in the user study, patients requiring surgery within 0-3 months are advised to seek an urgent referral to a glaucoma specialist. Those needing surgery within a 3-12 months time frame are recommended to schedule a referral with the glaucoma specialist in the near future (3-12 months). Meanwhile, patients not requiring surgery within a year do not need an appointment with a glaucoma specialist.

\paragraph{Study cases}
We selected 20 cases from a retrospective longitudinal study of glaucoma patients conducted at the Wilmer Eye Institute. The data corresponds to patients who had their first VF, OCT, and clinical ophthalmology assessment (baseline visit) on the same day. Surgical candidates were selected based on the criteria that the period from their baseline visit to their surgery fell within the previously defined time frames, such as within three months and between three months and one year, indicating more urgent cases. Similarly, non-surgical patients were included if the time between their baseline and second ophthalmology consultations fell within the specified time horizons, and they also had a follow-up consultation. 
Our case selection for the user study was as follows: 

\begin{itemize}
    \item Four eyes that underwent surgery within 0-3 months
    \item Four eyes that underwent surgery within 3-12 months
    \item Eight eyes that  did not undergo surgery within 12 months
     \item Four eyes that were inaccurately predicted by both the DLM and scorecard model (three false positives and one false negative, mirroring the DLM prediction performance)
\end{itemize}

To ensure that the model's performance does not confound the effects of different types of AI interventions, we specifically chose the cases where both the DLM and credit scorecard models provided identical predictions for all 20 cases. The only difference between the models is the explanation method used. To maintain a uniform difficulty across the four groups (Human only or baseline, No Explanation, Feature Importance, and Scoring-based explanations), we randomly select (without replacement) one eye from the 0-3 month interval, one from the 3-12 month interval, two eyes not needing surgery within a year, and one from the wrongly predicted category (misclassifying cases within 0-3 months and no surgery). This ensures that the accuracy of the AI recommendation in the No Explanation, Feature Importance Explanations, and Scoring-based Explanations stands at 80\%. In total, participants evaluated 20 cases that were distributed across the four experimental conditions following the strategy described above. The order of the groups with AI support and the order of the cases within each group were random. Each participant interacted with all four conditions consistent with a within-subjects design. 

\paragraph{Study Procedure}
In an online study, participants were first presented with a consent form and instructions. Then the participants went through the following steps: 1) They were asked demographics questions about age, years of experience as an optometrist, completion of optometry residency, familiarity with AI on a 5-point scale. To prevent the participation of individuals pretending to be optometrists, we implemented a screening process requiring participants to correctly answer a multiple-choice question, with a maximum of three attempts allowed. 2) In the main task, participants first assessed the patient vignettes under the Human Only condition. The subsequent three conditions included AI support and were presented in a random order. While the clinical data of 20 cases were fixed, the order and the AI explanations for the cases were randomized for each participant. Participants then make the final referral recommendation to decide if the patient needs a referral to glaucoma specialist for surgical intervention within 3 months, 3-12 months, or does not need a referral currently, along with the confidence level in their responses. 3) For each patient vignette under the groups with AI support, we further asked participants to rate their level of trust with the AI recommendation and the extent to which they integrated the AI recommendation in their decision-making process. 4) After evaluating the five patient vignettes, a post-group survey was administered to all experimental groups. This survey included questions about perceived effort and challenge (workload). Additionally, for groups receiving AI support, questions about the intention to use and the helpfulness of AI recommendations were also included.

\paragraph{Participants}

We recruited participants who are optometrists and >18 years old. We posted the survey on the American Academy of Optometry and various listservs within our institution. Participation in this study is completely voluntary and contingent upon agreement to the informed consent. To ensure data quality, we excluded data from participants whose median responding time was less than five seconds or those who provided the same recommendation to all cases. In total, 87 optometrists joined the study. Participants’ age, experience, completion of an optometry residency, proficiency with AI, and the preferred model evaluation are presented in Table~\ref{tab:sample}. Around half of the participants had optometry residency. The median years of experience of participants is 9 (5.0, 19.5). The majority of participants (> 80\%) indicated a high level of familiarity with AI . Participants who successfully completed the study entered  a lottery to receive one out of ten \$100 USD Amazon gift cards. The study was approved by the Johns Hopkins Medicine Internal Review Board. The study took 15 - 20 minutes on average to complete.

\begin{table}[]
\centering
\label{tab:sample}
\caption{Characteristics of the sample of 87 participants.}
\begin{tabular}{l|lll}
Numerical variables & \multicolumn{3}{c}{Median (IQR)} \\ \hline
Age (years) & \multicolumn{3}{c}{35 (32.0, 48.0)} \\
Years of experience & \multicolumn{3}{c}{9 (5.0, 19.5)} \\ \hline
Categorical variables & \multicolumn{1}{c|}{Group} & \multicolumn{1}{c|}{Count} & \multicolumn{1}{c}{\%} \\ \hline
\multirow{2}{*}{Optometry residency} & \multicolumn{1}{l|}{Yes} & \multicolumn{1}{l|}{45} & 51.7 \\
 & \multicolumn{1}{l|}{No} & \multicolumn{1}{l|}{42} & 48.3 \\
\multirow{5}{*}{AI familiarity} & \multicolumn{1}{l|}{Not familiar at all} & \multicolumn{1}{l|}{1} & 1.2 \\
 & \multicolumn{1}{l|}{Moderately unfamiliar} & \multicolumn{1}{l|}{2} & 2.3 \\
 & \multicolumn{1}{l|}{Neither unfamiliar nor familiar} & \multicolumn{1}{l|}{9} & 10.3 \\
 & \multicolumn{1}{l|}{Moderately familiar} & \multicolumn{1}{l|}{65} & 74.7 \\
 & \multicolumn{1}{l|}{Extremely familiar} & \multicolumn{1}{l|}{10} & 11.5
\end{tabular}
\end{table}

\paragraph{Measures}
We used a set of objective and subjective metrics to evaluate participants’ experience in the referral decision making task. We collected several objective metrics:

\begin{itemize}
    \item  \textit{Overall performance:} To assess participants' performance, we compared their answers to the ground truth of each patient case and computed the accuracy in each condition. If the participant's answer matched the decision for surgery on the clinical records, it was considered a correct referral. Conversely, if the participant's answer differed from the recorded clinical decision, it was considered an incorrect referral decision.

    \item \textit{Time:} This metric measures the duration in seconds that each participant spends reviewing the clinical data, considering the AI's recommendations (if provided), making the referral decision, and responding to the objective matrix.

    \item \textit{Agreement with the AI:} We quantified agreement as the percentage of cases where participants followed the AI suggestions in each condition that involved AI support. In order to assess users' willingness to adhere to the AI's recommendations, we compared participants' answers to the AI suggestions for each case. Agreement with the AI's recommendations was indicated when a participant's response matched the AI suggestion, while a difference in response signified disagreement with the AI's advice. We further analyze cases when participants accepted an AI’s incorrect prediction to measure overreliance~\cite{vasconcelos2023explanations}. 

    \item \textit{Adjustments to AI predictions:} presenting participants directly with AI predictions can lead to anchoring bias as they adjust insufficiently from the anchor, i.e., the AI prediction45. Average disagreement is commonly computed to quantify adjustment from AI predictions when the decision task has binary outcome, but with the potential outcomes of referral recommendations, we are interested in a finer grained analysis. We measure the extent of participant deviations from AI predictions by computing the difference between the AI-suggested referral category and the participant's final response. We define the AI deviation score as a numerical variable, where zero indicates no adjustment, or agreement with the AI prediction. Disagreement between AI suggestions and participant responses can occur across one or two referral categories. For differences spanning one category (deviation=1), scenarios include: i) AI: no referral is currently needed, participant: referral within 3-12 months, ii) AI: referral within 3-12 months, participant: no referral is currently needed, iii) AI: referral within 3 months, participant: referral within 3-12 months, and iv) AI: referral within 3-12 months, participant: referral within 3 months. For differences spanning two categories (deviation=2), the discrepancy is more pronounced: i) AI: referral within 3 months, participant: no referral is currently needed and ii) AI: no referral is currently needed, participant: referral within 3 months. 

    We did not consider the direction of the changes, i.e., whether final responses result in an increase or decrease of the urgency level to be seen by a specialist. Smaller deviations mean participants adjusted less their final responses, and could have been more anchored to the AI prediction. 

    \item \textit{Over-claims:} We identified instances where participants incorrectly increased the perceived urgency of a referral or suggest a referral when none is warranted, and report over-claims as a percentage of such decisions among all cases susceptible to over-claims, i.e., cases that need a referral within 3-12 months and cases that do not need a referral currently. This includes cases where "The patient doesn't need a referral currently" is misclassified as "The patient needs a referral to a glaucoma specialist within 3 months" or "The patient needs a referral to a glaucoma specialist within 3-12 months", and cases where "The patient needs a referral to a glaucoma specialist within 3-12 months" is misclassified as "The patient needs a referral to a glaucoma specialist within 3 months".

    \item \textit{Under-claims:} We identified instances where participants incorrectly diminish the perceived urgency of a referral or do not suggest a referral at all when needed, and report under-claims as a percentage of such decisions among all cases susceptible to under-claims, i.e., cases that need a referral within 3 months and cases that need a referral within 3-12 months. This includes cases where "The patient needs a referral to a glaucoma specialist within 3 months" is misclassified as "The patient needs a referral to a glaucoma specialist within 3-12 months" or "The patient doesn't need a referral currently", and cases where "The patient needs a referral to a glaucoma specialist within 3-12 months" is misclassified as "The patient doesn't need a referral currently".

\end{itemize}

In addition, we collected several subjective metrics using a five-point Likert scale from 1 (Strongly disagree) to 5 (Strongly agree). Participants rated statements regarding confidence in their responses and workload in terms of effort and frustration~\cite{cai2019human}. In the groups using AI assistance, participants rated their agreement with the previous statements and also statements regarding trust in the AI’s suggestions, integration of AI suggestions into their decision making process, helpfulness of the AI’s suggestions, and future use of the AI-based system. The complete set of statements is included in the appendix.

\section{Results}
For the analysis of results, we first employed one-way repeated measures Analysis of Variance (ANOVA) tests on continuous objective metrics (accuracy, time, agreement rate, referral errors, AI deviation score) and Aligned Rank Transform (ART) ANOVA tests on subjective metrics (including confidence, workload, perceived challenge, trustworthiness, decision-making, helpfulness, and willingness for clinical practice). In the models, we define the type of decision support as the within subjects fixed effect and participant ID as the random effect. We validated normality and homogeneity of variance assumptions using the Shapiro-Wilk and Levene’s tests, and when violated used Friedman’s test, the non-parametric alternative for repeated measures ANOVA. For measurements with multiple observations within each experimental group, e.g., task time, we employed the non-parametric ART ANOVA. If the results have statistical significance, we further performed post-hoc pairwise statistical comparisons between the two groups using paired t-tests or Wilcoxon tests with Holm method correction. For all the statistical tests reported below, significance levels below .05 were considered as a statistically significant effect. We report effect sizes using partial eta squared for ANOVA models, where 0.01 is considered small, 0.06 moderate, and 0.14 a large effect. When Friedman’s test is used, we report Kendall’s W coefficient, where 0.1-0.3 is considered small, 0.3-0.5 moderate, and greater than 0.5 a large effect. 

\begin{figure}
    \centering
    \includegraphics[width=0.80\linewidth]{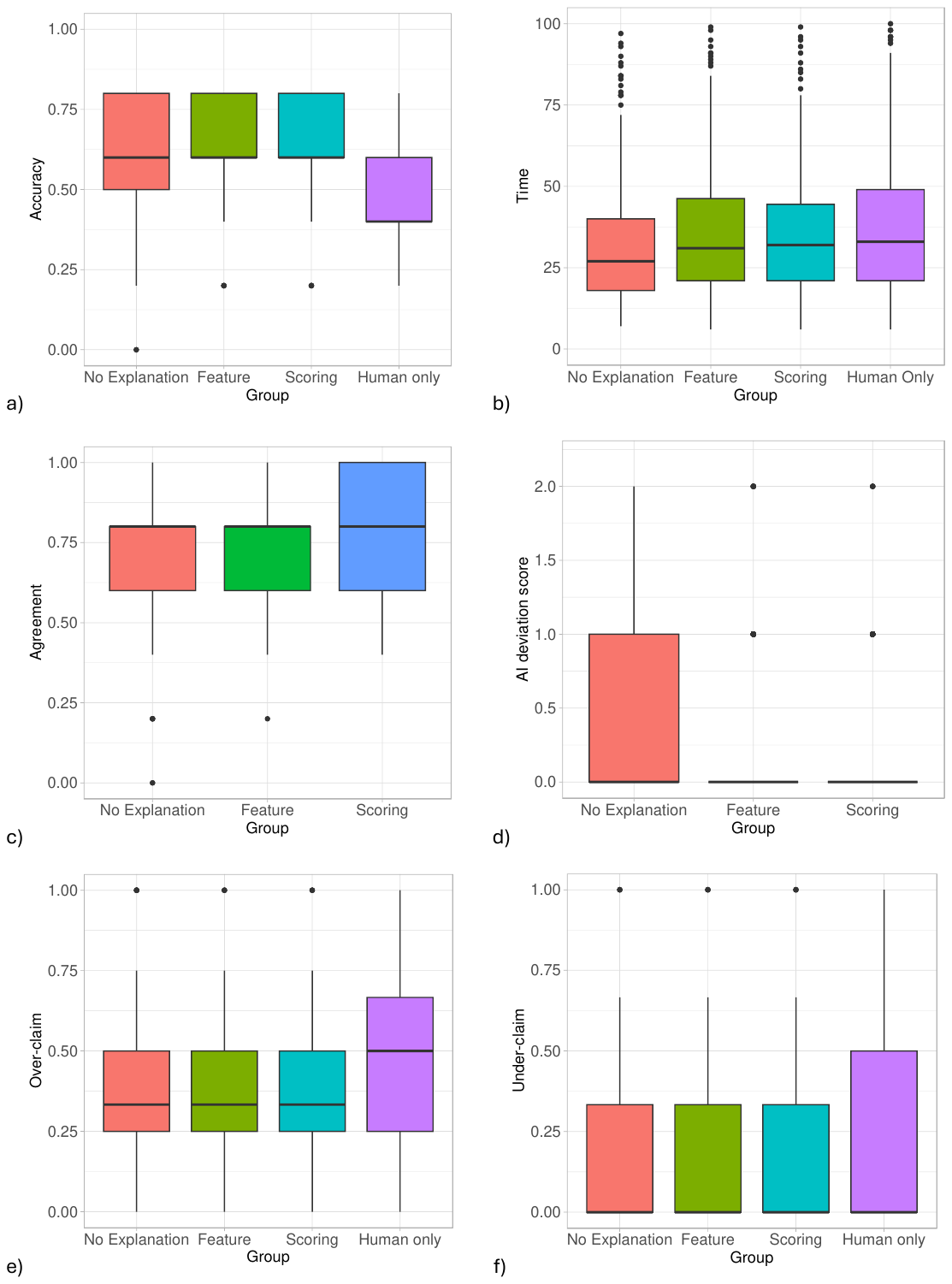}
    \caption{Box plots for objective measures across different explanation conditions and baselines. Agreement and AI deviation scores only include groups with AI support.}
    \label{fig:results_obj}
\end{figure}

\begin{figure}
    \centering
    \includegraphics[width=1\linewidth]{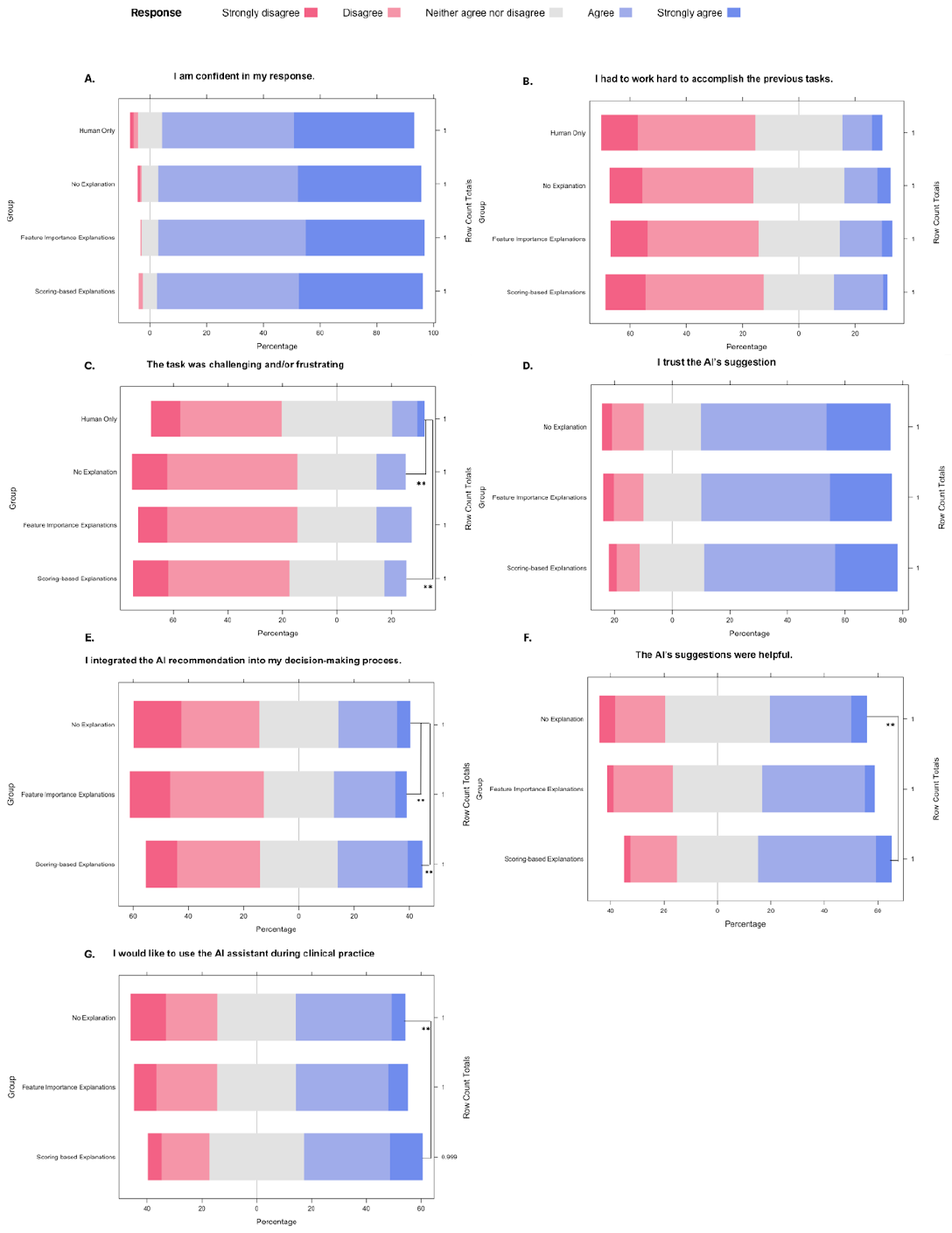}
    \caption{Distribution of five-point rating scale responses in for the following constructs: A) Confidence, B) Workload - Effort, C) Workload - Frustration, D) Trust, E) Support decision making, F) Helpfulness, and G) Future use. For each measure, explanation groups (No Explanation, Feature Importance, Scoring-based) and baseline (on the top if present) are shown.}
    \label{fig:subjective}
\end{figure}

In summary, the evidence from our study suggests that AI-based systems can assist optometrists in achieving more accurate referrals, while also contributing to a reduction in participants' perceived levels of challenge and frustration. However, referral's accuracy was higher when the AI model operates alone. AI support without explanations has a positive effect on reducing the time required for referral decision-making and adjusting away from AI predictions. Feature Importance Explanations positively influence participants in integrating AI into the decision-making process, but may cause overreliance. Scoring-based Explanations positively impacts perceived helpfulness, decreases overtreatment rate, higher perceived integration of AI into decision-making, and willingness for deployment in clinical practice.
Table~\ref{tab:objective} aggregates the descriptive statistics of participants' performance, decision time, and outcomes of referral decisions for each condition. 
Figures~\ref{fig:results_obj} and ~\ref{fig:subjective} present data distributions of objective and subjective measures, respectively.

\begin{table}[]
\label{tab:objective}
\caption{Descriptive statistics including mean (standard deviation) and 95\% confidence interval in (1) Human Only: Control group – Referral Decision-Making without a Prediction Model; (2) No Explanation: AI prediction provided without explanation; (3) Feature Importance: AI prediction provided with Top 3 most important features; (4) Scoring-based: AI prediction presented with a Scorecard explanation.}
\begin{tabular}{l|llll}
Measure & Human Only & No Explanation & \begin{tabular}[c]{@{}l@{}}Feature \\ Importance\end{tabular} & Score-based \\ \hline
Accuracy & \begin{tabular}[c]{@{}l@{}}0.51 (0.16)\\ {[}0.46, 0.56{]}\end{tabular} & \begin{tabular}[c]{@{}l@{}}0.58 (0.22)\\ {[}0.56, 0.60{]}\end{tabular} & \begin{tabular}[c]{@{}l@{}}0.60 (0.16)\\ {[}0.55, 0.64{]}\end{tabular} & \begin{tabular}[c]{@{}l@{}}0.62 (0.17)\\ {[}0.57, 0.66{]}\end{tabular} \\
Time & \begin{tabular}[c]{@{}l@{}}35\\ {[}22, 57{]}\end{tabular} & \begin{tabular}[c]{@{}l@{}}28\\ {[}18, 43{]}\end{tabular} & \begin{tabular}[c]{@{}l@{}}33\\ {[}22, 50{]}\end{tabular} & \begin{tabular}[c]{@{}l@{}}33\\ {[}22, 50{]}\end{tabular} \\
Agreement rate & - & \begin{tabular}[c]{@{}l@{}}0.72 (0.24)\\ {[}0.67, 0.77{]}\end{tabular} & \begin{tabular}[c]{@{}l@{}}0.78 (0.18)\\ {[}0.74, 0.82{]}\end{tabular} & \begin{tabular}[c]{@{}l@{}}0.79 (0.18)\\ {[}0.75, 0.83{]}\end{tabular} \\
Overreliance & - & \begin{tabular}[c]{@{}l@{}}0.77 (0.42)\\ {[}0.68, 0.86{]}\end{tabular} & \begin{tabular}[c]{@{}l@{}}0.92 (0.27)\\ {[}0.86, 0.98{]}\end{tabular} & \begin{tabular}[c]{@{}l@{}}0.85 (0.36)\\ {[}0.78, 0.93{]}\end{tabular} \\
Over-claims rate & \begin{tabular}[c]{@{}l@{}}0.49 (0.25)\\ {[}0.44, 0.55{]}\end{tabular} & \begin{tabular}[c]{@{}l@{}}0.42 (0.24)\\ {[}0.37, 0.47{]}\end{tabular} & \begin{tabular}[c]{@{}l@{}}0.40 (0.25)\\ {[}0.35, 0.45{]}\end{tabular} & \begin{tabular}[c]{@{}l@{}}0.38 (0.23)\\ {[}0.33, 0.43{]}\end{tabular} \\
Under-claims rate & \begin{tabular}[c]{@{}l@{}}0.27 (0.33)\\ {[}0.20, 0.34{]}\end{tabular} & \begin{tabular}[c]{@{}l@{}}0.20 (0.27)\\ {[}0.14, 0.26{]}\end{tabular} & \begin{tabular}[c]{@{}l@{}}0.19 (0.26)\\ {[}0.14, 0.25{]}\end{tabular} & \begin{tabular}[c]{@{}l@{}}0.19 (0.25)\\ {[}0.14, 0.25{]}\end{tabular} \\
AI deviation score & - & \begin{tabular}[c]{@{}l@{}}0.33 (0.57)\\ {[}0.28, 0.39{]}\end{tabular} & \begin{tabular}[c]{@{}l@{}}0.24 (0.46)\\ {[}0.20, 0.28{]}\end{tabular} & \begin{tabular}[c]{@{}l@{}}0.22 (0.44)\\ {[}0.18, 0.26{]}\end{tabular}
\end{tabular}
\end{table}

\paragraph{How do AI-based systems and their explanations impact the quality of participants’ referral decisions and influence them?}

\paragraph{Accuracy}

The model’s evaluation in our test set showed that the scorecard model has an average accuracy across the three time horizon categories of 0.77 and 0.71 for the black box and scorecard models, respectively. Additional performance metrics are presented in Table~\ref{tab:metrics}. Based on the selection of patient cases for the user study, the AI’s accuracy in this subset of patients was 0.80 (only one incorrect recommendation in the five patients per experimental group). We observed significant differences in the accuracy across different conditions using Friedman’s test ($\chi^2(3)=22.3, p<.001, W=0.09$). In the baseline condition, participants alone achieved a mean accuracy of 0.51 (95\% CI: 0.46, 0.56). In the presence of Scoring-based Explanations, accuracy was on average 0.62 (95\% CI: 0.57, 0.66), surpassing both 0.58 in No Explanation (95\% CI: 0.57, 0.66) and 0.60 (95\% CI: 0.55, 0.64) in Feature Importance Explanations. Post-hoc Wilcoxon tests with the Holm correction method revealed that participants in the AI-based groups had a significantly higher accuracy compared to the baseline condition (No Explanation: $p=.005$, Feature Importance: $p=.001$, Scoring-based: $p<.001$). Participants in the AI-based groups achieved better performance, but there were no statistically significant differences between the No Explanation, Feature Importance Explanations, and Scoring-based Explanations conditions.

\begin{table}[]
\label{tab:metrics}
\caption{Performance metrics of the black box DNN and intrinsically explainable Scorecard model on the separate test set. Each metric is reported by category and the average.}
\begin{tabular}{l|cccc|cccl}
Metric & \multicolumn{4}{c|}{Precison} & \multicolumn{4}{c}{Recall} \\ \hline
Category & No surgery & 0-3M & \multicolumn{1}{c}{3-12M} & Avg & No surgery & 0-3M & \multicolumn{1}{c}{3-12M} & \multicolumn{1}{c}{Avg} \\ \hline
DNN & 0.89 & 0.52 & \multicolumn{1}{c}{0.17} & 0.53 & 0.92 & 0.35 & \multicolumn{1}{c}{0.23} & 0.50 \\
Scorecard & 0.93 & 0.56 & \multicolumn{1}{c}{0.12} & 0.54 & 0.76 & 0.69 & \multicolumn{1}{c}{0.25} & 0.57
\end{tabular}
\end{table}

\paragraph{Time}
We compared the time required for participants to complete the referral task across all the experimental groups. Because the normality assumption was not met ($p<.05$), we used an ART ANOVA to measure the effect of the type of support on task time. We found a significant difference in the average time that participants spend based on the type of support they receive ($F(3,258)=10.57, p<.001, \eta^2=0.11$). Post-hoc pairwise comparisons using the Holm-Bonferroni correction indicated that when participants received AI predictions only in the No Explanation group (median=28, IQR=[18, 43]), they spent significantly less time compared to the non-assisted group (median=35, IQR=[22, 57]), and also when comparing against the Feature Importance (median=33, IQR=[22, 50]) and Scoring-based Explanations (median=33, IQR=[22, 50]) groups (all comparisons <.001). No significant differences were observed between the latter two and the human Only group or between the Feature Importance and Scoring-based Explanations. 

\paragraph{Agreement with AI predictions}
We evaluated whether participants’ agreement with AI predictions was affected by the type of explanations provided using Friedman's test since the normality assumption was violated ($p<.05$). We observed that there is no significant difference in agreement with AI recommendations across different types of AI support ($\chi^2(2)=5.88, p=.053$). Within the Scoring-based Explanations group, the average agreement with AI suggestions was the highest with 0.79 (95\% CI: 0.77, 0.80), followed by group Feature Importance Explanations 0.78 (95\% CI: 0.76, 0.79) and No Explanation 0.72 (95\% CI: 0.70, 0.74).

Because all AI models had an accuracy of 80\%, we further analyze agreement with incorrect AI predictions to measure potential overreliance. The normality (p<.001) and variance (p<.05) assumptions were violated and Friedman’s test was used to measure whether the type of AI support affects overreliance. We found that overreliance is significantly different across AI support types ($\chi^2(2)=10.2, p=.006, W=0.06$). Post-Hoc Wilcoxon tests showed significantly higher ($p=.02$) agreement when AI predictions are presented with Feature Importance explanations ($M=0.92, SD=0.27$) compared to not presenting explanations at all ($M=0.77, SD=0.42$). No significant differences were observed with respect to the Scoring-based Explanations ($M=0.85, SD=0.36$).

\paragraph{Over-claims and Under-claims}
The AI errors could either be under-claims (i.e. 0-3 months cases predicted as no surgery) or over-claims (i.e. no surgery cases predicted as within 0-3 months). In general, underclaims were less common as a type of AI error (occurred in around 30\%). 
We observed significant differences in the fraction of over-claims across different conditions ($F(3,258)=5.68, p <0.001, \eta^2=0.06$). Post-hoc pairwise comparisons using the Holm-Bonferroni correction revealed that the over-claims rate in the Scoring-based Explanations group ($M=0.38, SD=0.23$) was significantly lower ($p=.013$) compared to the Human Only condition ($M=0.49, SD=0.25$). The fraction of over-claims was on average 0.42 ($SD=0.24$) and 0.40 ($SD=0.25$) in the No Explanation and Feature Importance groups, respectively. No further differences were found in the post-hoc comparisons. 

Similarly, we measured the effect of the type of support on the fraction of under-claims. Since the normality assumption was violated ($p<.001$), we used a Friedman’s test. The test revealed no significant difference in under-claims across the conditions with AI support nor with the Human Only condition ($\chi^2(3)=3.76, p=.288$). In the Human Only group, the average under-claims fraction was 0.27 ($SD=0.33$). Under-claims fraction was similar across all the groups that used AI support (No Explanation: $M=0.20, SD=0.27$, Feature Importance: $M=0.19, SD=0.26$, Scoring-based: $M=0.19, SD=0.25$). 

\paragraph{Adjustments to AI predictions}
We assessed the impact of different types of AI support on the discrepancies between participants’ final responses and AI predictions by analyzing the AI deviation score. Because both the normality (p<.001) and variance ($p<.05$) assumptions were violated, we used the non-parametric alternative ART ANOVA. We found a significant effect of the type of AI support on the AI deviation score ($F(2,172)=4.85, p=.008, \eta^2=0.05$). The average AI deviation score in the No Explanation group was 0.33 ($SD=0.57$), while the average scores in the Feature Importance and Scoring-based Explanations were 0.24 ($SD=0.46$) and 0.22 ($SD=0.44$), respectively. Post-hoc comparisons indicated significantly larger deviations in participants’ responses in the No Explanation group compared to the Feature Importance ($p=.030$) and Scoring-based Explanations ($p=.013$) groups. No significant differences were found between the two types of explanations. 

\paragraph{How do AI-based systems and their explanations affect participants’ perceptions during referrals decision making?}

\paragraph{Perceived challenge to complete the task}
We observed significant differences in participants’ ratings of the perceived challenging and/or frustrating level across different groups ($F(3,258)= 3.48, p=.017, \eta^2=0.04$). Post-hoc pairwise comparisons using the Holm-Bonferroni correction indicated that participants' perceived level of challenge and frustration in completing the referral task was significantly higher in the Human Only group ($M=2.61, SD=0.87$) compared to the Scoring-based Explanations group ($M=2.38, SD=0.81, p=.035$) and No Explanation group ($M=2.38, SD=0.84, p=.027$). However, no significant differences were observed with Feature Importance explanations ($M=2.46, SD=0.86$) or between the different AI-based systems. 

\paragraph{Integration of AI Recommendations}
We observed significant differences in participants’ self-reported integration of AI recommendations into their decision making processes across different AI-based conditions ($F(2,172)=4.58, p=.012,  \eta^2=0.05$). Post-hoc pairwise comparisons using the Holm-Bonferroni correction indicated that participants’ perception of integrating AI recommendations into their decision-making was significantly higher in the Scoring-based Explanations group ($M=2.83, SD=1.09$) compared to the No Explanation ($p=.027$) and Feature Importance Explanations groups ($p=.027$). No significant differences were found between the No Explanation ($M=2.67, SD=1.13$) and Feature Importance Explanation groups ($M=2.67, SD=1.10$). 

\paragraph{Helpfulness of the AI suggestions}
We observed significant differences in participants' ratings of the usefulness of the AI's suggestions across different AI-based conditions ($F(2,172)=3.85, p=.023,  \eta^2=0.04$). Post-hoc pairwise comparisons using the Holm-Bonferroni correction indicated that usefulness ratings were significantly higher ($p=.023$) in the Scoring-based Explanations group ($M=3.34, SD=0.91$) compared to the No Explanation group ($M=3.13, SD=0.97$), while no significant differences were observed against the Feature Importance Explanations group ($M=3.20, SD=0.90$). Likewise, no significant differences were observed between the No explanation and Feature Importance Explanation groups. 

\paragraph{Willingness for deployment in clinical practice}
We observed significant differences in participants' ratings of their willingness to deploy AI in clinical practice across different AI-based conditions ($F(2,172)=4.58, p=.012, \eta^2=0.05$). Post-hoc pairwise comparisons using the Holm-Bonferroni correction indicated that participants' willingness to deploy AI in clinical practice was significantly higher ($p=.037$) in the Scoring-based Explanations group ($M=3.28, SD=1.03$) compared to the No Explanation group ($M=3.01, SD=1.12$), while no significant differences were observed compared to the Feature Importance Explanations group ($M=3.09, SD=1.07$). Likewise, no significant differences were observed between the No explanation and Feature Importance Explanation groups. 

The majority of the participants preferred the Scoring-based Explanations (54\%), followed by the Feature Importance (31\%), and working on their own, as in the Human Only group (12.7). Only 2.3\% preferred AI support without explanations.

\paragraph{Trust, confidence, and effort}
We observed that there is no significant difference in participants' ratings of trust in the AI recommendations across different AI-based conditions ($F(2,172)=0.11, p=.895$). There were no significant differences in participants’ confidence in their referral decisions ($F(3,258)=0.62, p=.602$) or in their perceived effort required to complete the referral task across the experimental groups ($F(3,258)=0.70, p=.556$).

\section{Discussion}
We conducted a study to assess the effects of AI support for identifying high-risk glaucoma patients. We compared the following scenarios: black box models that only provide recommendations, transparent models that offer explanations post-hoc or directly show how inputs are transformed into outputs, and a control group without any type of AI support. This structure allowed us to evaluate how different types of explanations influence participants’ experience and performance. We built these algorithms to provide more realistic predictions to the patient vignettes and the presentation of feasible explanations for such cases. 

Consistent with prior research~\cite{lai2019human,buccinca2020proxy}, our results demonstrate that the Human-AI team performed better overall than participants who completed the task without any AI support, but still did not surpass the AI model alone (accuracy was set to 80\% during the study). These results support hypothesis H1a. However, there were no statistically significant differences across the different types of AI support, suggesting a lack of evidence for hypothesis H1b. Our findings challenge the widely held assumption that providing explanations alongside AI suggestions will markedly improve clinicians' performance, echoing similar observations in prior empirical evaluations of AI systems in healthcare~\cite{chanda2024dermatologist}. A potential reason is that explanations may not help participants to recognize incorrect recommendations or situations in which following the model advice increases the chances of reaching the correct decision~\cite{naiseh2023different}. Our findings indicate a tendency toward increased overreliance on incorrect AI recommendations when post-hoc explanations are provided, compared to when AI operates as a black box. However, given the small effect size and the constrained sample—each group received only one incorrect AI prediction—these results should be interpreted with caution. Explanations alone may not be sufficient to improve outcomes, suggesting the need for a deeper understanding of how clinicians interact with and interpret AI-generated information. 

Analyzing task efficiency, we found that participants in the No Explanation condition spent on average significantly less time completing the task compared to the unassisted group and both groups with explanations. There was no statistically significant difference in the time spent by participants between the Human Only, Feature Importance Explanations, and Scoring-based Explanations conditions. We attribute this finding to the fact that Feature Importance Explanations and Scoring-based Explanations complement referral recommendations by providing more text and visual information for the user to process, which may have contributed to spending a similar amount of time compared to the Human Only condition. Along with the performance improvements, we notice that AI support with No Explanation can improve both the efficiency and effectiveness of the decision-making process. With a more thorough analysis of inaccurate referral decisions, we noticed that the rate of over-claims of the severity of the cases is significantly lower in the Scoring-based Explanations compared to the Human Only condition, but not different across different types of AI support. This disparity suggests a potential for the supplementary information provided by intrinsic explanations to assist in minimizing instances of unnecessary treatments, also referred to as false positives. In our study, we found no differences or advantages associated with any type of AI support in terms of underestimating the severity of a case. Reducing these instances is crucial, as they represent false negative cases, i.e., patients who require timely interventions but do not receive it. 

Although providing AI support did not significantly increase participants' confidence in their referral decisions or decrease workload, certain types of AI support did significantly decrease participants' perceived levels of challenge and frustration to complete the referral task. AI predictions without explanations and in the Scoring-based model helped to reduce the level of challenge and frustration compared to the unassisted condition, but neither type of explanation presented an advantage. Therefore, our findings partially support hypothesis H2a. Presenting AI suggestions with explanations (vs. not presenting explanations at all) did not significantly increase participants' confidence level, trust, or decrease participants' perceived levels of effort and challenge or frustration experienced during the task. Nevertheless, participants' ratings of usefulness and future use in clinical practice for the Scoring-based Explanations were significantly higher compared to using AI predictions without explanations. To illustrate the usefulness of the Scoring-based Explanations, one participant claimed: “I thought that having all of the factors listed together was very helpful, and that the visual of scale as to where the patient fell in reference to the scores were helpful.” These results partially support hypothesis H2b. In addition, more than half of the participants preferred Scoring-based Explanations. Consistent with the previous research, the subjective evaluations of explainable systems, such as measures of trust and preference, may not directly correspond to the behavior and success of participants’ interaction with AI support~\cite{buccinca2020proxy}. One possible explanation for this is that all the AI recommendations have the same accuracy but the transparent rules and comprehensive information sharing, along with the graphical representation of a numerical assessment in the Scoring-based Explanations, help participants understand the reasoning behind the predictions. Additionally, participants' familiarity with scorecard systems, widely used in the credit system and healthcare, may have contributed to their increased willingness to use the AI system with Scoring-based Explanations during clinical practice.

Another interesting finding is that while participants reported varying levels of use of AI recommendations from different model types—with greater use reported for intrinsically interpretable models—the observed agreement rate was considerable (greater than 76\%), but did not change significantly with the presence or type of explainability incorporated. Similar agreements with AI predictions regardless of explanations were reflected in similar trust levels reported for AI predictions in the No Explanation, Feature Importance, and Scoring-based explanation. Despite the similar agreement rate with AI recommendations across groups with AI support, our analysis revealed variations in how participants deviated from these recommendations. Participants displayed larger deviations in their responses with the black box model compared to both the Feature Importance and Scoring-based explanations. Smaller average AI deviation scores, driven largely by the prevalence of zero values from multiple cases of agreement, may suggest an anchoring effect to AI predictions, which may limit participants' exploration of alternative referral options. We found no evidence that Feature Importance or Scoring-based explanations result in greater or lesser anchoring compared to each other. Anchoring itself is not inherently negative; adhering to recommendations from a highly accurate source can indeed enhance the outcomes of human-AI collaboration. However, it is crucial to ensure that participants remain conscious of their decisions, underscoring the importance of investigating anchoring biases in AI as decision support.

\paragraph{Limitations and future work}
Our work has several limitations. First, the measurements of agreement and decision adjustments may have some constraints. We compared the participants' final referral decision to the AI recommendation. The design of the experiment does not facilitate precise measurement of the impact of AI advice on decision-making, as decisions were made concurrently with the presentation of AI suggestions. A more effective approach might be to directly inquire whether participants agree with the suggestions offered by the AI system, thus providing a clearer and more direct assessment of the advice's influence~\cite{gomez2023mitigating}. Alternatively, in a two-step workflow, participants are presented with AI inferences only after they had made an initial task assessment~\cite{fogliato2022goes}. Additionally, it is crucial to acknowledge that our study's results and conclusions may reflect the perspectives and experiences of a specific group—optometrists, who constituted our participant pool. Therefore, caution should be exercised when generalizing these findings to other populations or contexts, considering the small effect sizes for some significant findings. Further validations with a larger sample size may enable further analysis on the effects of different interaction paradigms when working with different AI explanations. 

Based on participants' feedback, it is important to consider topical medications, systemic medications, peak pressure, and goal pressure when referring to a glaucoma specialist. In future work, we plan to improve the DLM model, adjust the user interface design to better integrate with optometrists' workflow, and provide additional information to them, such as historic  VF/RNFL/IOP measurements. Furthermore, involving participants in the decision-making process and giving them an active role could potentially enhance performance by properly identifying AI failure models and improve the user experience~\cite{gomez2023mitigating}.

\section{Conclusion}
Our investigation into the use of AI to support primary eye care providers in making referral decisions for glaucoma patients provides critical insights. While our explainable AI algorithms for surgical intervention prediction showcase the technical feasibility and potential to identify high-risk patients, the actual integration of these tools into clinical decisions highlighted both promises and challenges. Our findings suggest that AI support can enhance the ability of non-specialists to identify high-risk patients requiring urgent referrals while simultaneously alleviating the perceived challenges of the task. However, bridging the gap to reach the algorithmic performance levels achieved by the AI alone remains a critical challenge as keeping the human-in-the-loop is critical in this context. Although our results did not show that explanations improve providers' performance by making the AI's decision-making process more transparent, they reveal an opportunity to enhance the collaborative Human-AI  decision-making process. Moving forward, the focus should be on developing AI tools that support both clinical effectiveness and user engagement, thereby maximizing the benefits of AI in healthcare.

\bibliographystyle{plainnat}
\bibliography{references}





\newpage


\end{document}